\documentclass[useAMS,usenatbib]{mn2e}
\usepackage{epsfig}

\begin{document}

\title{Role of thermal conduction in an advective  accretion  with bipolar outflows}

\author[F. Khajenabi \& M. Shadmehri]{Fazeleh Khajenabi\thanks{E-mail:
f.khajenabi@gu.ac.ir;} and Mohsen Shadmehri\thanks{Email:m.shadmehri@gu.ac.ir}\\
School of Physics, Faculty of Science, Golestan University, Gorgan 49138-15739, Iran}

\maketitle

\date{Received ______________ / Accepted _________________ }

\begin{abstract}
Steady-state advective accretion flows in the presence of thermal conduction are studied. All three components of velocity in a spherical coordinates are considered and   the flow displays both inflowing and outflowing regions according to our similarity solutions. Thermal conductivity provides latitudinal energy transport and so,  the flow rotates more slowly and becomes hotter with increasing  thermal conductivity coefficient. We also show that opening angle of the outflow region decreases as thermal conduction becomes stronger.

\end{abstract}

\begin{keywords}
galaxies: active - black hole: physics - accretion discs
\end{keywords}
\section{Introduction}
The importance of the advection of  energy in the accreting systems has been recognized  by \cite{Ichi} and  \cite{narayan94,narayan95}. In this type of  accretion, which is  known as the Advection-Dominated Accretion Flow (ADAF), the generated heat due to the turbulence is advected with the flow instead of immediately radiating out of the system. Although the original ADAF model is simply presented by a set of time-independent self-similar solutions \citep{narayan94}, subsequent studies confirmed validity of the similarity solutions by studying  steady-state global solutions  \citep[e.g.,][]{narayan97,chen}. Most of the previous analytical ADAF solutions are either based on the height-averaged equations in the cylindrical coordinates or in the spherical coordinates. It seems for describing ADAF accretion, however, spherical system is more adequate simply because ADAFs are generally hot and geometrically thick.

On the other hand, the ADAF solutions are convectively unstable \citep{narayan94}, and so, another type of the accretion has been suggested, i.e. Convection-Dominated Accretion Flow \citep[e.g.,][]{Igu,bland2004,zhang}. Moreover, the original ADAF solution imply a positive value for the Bernoulli parameter. This point motivated \cite{bland} to present solutions with outflows \citep[also see,][]{begelman2012}.  These kinds of the models for the accretion flows are  classified  as the Radiatively Inefficient Accretion Flows (RIAFs). Over recent decade, this type of accretion has been extended in the various directions, such as considering the role of the magnetic fields \citep[e.g.,][]{Aki},  outflows \citep[e.g.,][]{Xue, Xu, shad08}, and contribution of the thermal conduction \citep[e.g.,][]{tanaka,John}. Recent numerical simulations of the accretion flows have also shown the existence of outflows \citep*[][]{feng1,feng2,narayan12}. \cite*{feng1} performed numerical simulations of hot accretion flows and found the accretion rate and the density are power law functions of the radial distance \citep*[see also,][]{feng2}. \cite{narayan12} have also presented hydrodynamical simulations of hot accretion flows but considering relativistic effects. However,  they discuss  outflows in their simulations are not as strong as previous studies. However, none of the recent numerical simulations have considered thermal conduction.

The mean free path of the particles in the accreting systems which are described by the ADAF solutions is generally larger than the typical size of the system \citep{John}. Thus, the system is weakly collisional and instead of applying the standard magnetohydrodynamic equations one should use weakly collisional equations. Under such circumstances, in particular, thermal conduction takes a saturated form where the heat flux due to the conduction is not proportional to the gradient of the temperature \citep[e.g.,][]{John}.

Some authors investigated  role of a saturated form of the thermal conduction in ADAFs using height-integrated equations \citep[e.g.,][]{shah,ghanbari09,fagh12a,fagh13,fagh12a}. Despite  the positivity of Bernoulli parameter for these one-dimensional ADAF solutions and  possible  emergence of bipolar outflows, it is unlikely to  study  one-dimensional ADAFs with outflows self-consistently unless the outflows are included in a phenomenological way \citep[e.g.,][]{shad08,bu}. Motivated by this fact, \cite{tanaka}  constructed two-dimensional axisymmetrical model of ADAFs with thermal conduction and showed that because of the latitudinal energy transport by the thermal conduction, the hot accretion flow spontaneously develops bipolar outflows. However, \cite{tanaka} neglected the latitudinal component of the velocity as \cite{narayan95} did and so, the accretion rate is independent of the radius despite of the existence of outflows. If the latitudinal component of the velocity is neglected, the continuity equation implies the accretion rate becomes independent of the radius. Under this condition, \cite{tanaka} found solutions with positive radial velocity near the axis and considered such solutions as outflows.   In order to study the two-dimensional structure of ADAFs with outflows, just a few authors considered all three components of the gas velocity  in a spherical coordinates  \citep{Xu,Xue,jiao}. But thermal conductivity has been neglected in all these works, despite the important role of the energy transport by the thermal conduction in generating outflows from the ADAFs. In this work, we address this issue by constructing steady-state two-dimensional models for ADAFs including all three components of the velocity and thermal conductivity. Basic assumptions and general equations are presented in the next section. Self-similar solutions will be analyzed in section 3. We will conclude by a summary of the results in section 4.

\section{general formulation}
Our basic equations are the standard hydrodynamics equations in the spherical coordinates $(r,\theta,\phi)$ where the central mass $M$ is at its origin. The main equations can be written as
\begin{equation}
\nabla.(\rho {\bf v}) =0,
\end{equation}
\begin{equation}
\rho {\bf v}.\nabla {\bf v} = -\nabla p - \rho\nabla\Phi + \nabla . {\bf \Pi},
\end{equation}
\begin{equation}
\rho ({\bf v}.\nabla e - \frac{p}{\rho^2} {\bf v}.\nabla\rho ) = f\nabla {\bf v} : {\bf \Pi} + \nabla . (\lambda \nabla T),
\end{equation}
where $\rho$, ${\bf v}$, $p$, ${\bf \Pi}$, $e$, $T$ and $f$ are density, velocity vector, gas pressure, tensor of viscous stress, internal energy of gas, gas temperature, and advective fraction of the dissipated energy. We use the Newtonian gravitational potential, $\Phi =-GM/r$. Also, $\lambda$ is the thermal conductivity coefficient.

Like  previous semi-analytical studies of the structure of ADAFs, however, it is assumed that the system is  steady-state and axisymmetric which imply that all  physical variables are independent of the time $t$ and the azimuthal angle $\phi$. Such a steady-state configuration can be studied using similarity technique, by which the latitudinal dependence and  the radial dependence of the variables are separated by assuming the radial part as a power-law function of the radius. For simplicity, the latitudinal component of the velocity (i.e., $v_{\theta}$) has been neglected in most of the previous studies except a few authors who studied dynamical role of outflows in the ADAFs.

Components of the equations of motion are complicated, in particular because of the presence of the components of the stress tensor. Interestingly, \cite{jiao} showed that one can reproduce the standard ADAF solutions \citep{narayan95},  if  the $r\phi$ component of the stress tensor is assumed to be dominant. This assumption not only simplifies the basic equations, but there is no need to impose boundary condition at the pole. \cite{jiao} also relaxed $v_{\theta}=0$ and obtained solutions with a field of inflow near the equatorial plane and outflow near the axis.  We also assume that the $r\phi$ component of stress tensor is dominant and $\alpha$ prescription is used, i.e. $\Pi_{r\phi}=-\alpha p$. As for the boundary conditions,  mirror symmetric condition at the equator is applied. We discuss below, however, in the presence of thermal conduction a boundary condition at the opening angle of outflow is needed in order to obtain a unique solution for a given set of the input parameters.

\section{self-similar solutions}
The following separable solutions are introduced,
\begin{equation}
\rho(r,\theta ) = \rho(\theta) r^{-n},
\end{equation}
\begin{equation}
v_r (r, \theta ) = v_r (\theta) \sqrt{GM/r},
\end{equation}
\begin{equation}
v_{\theta}(r,\theta ) = v_{\theta}(\theta ) \sqrt{GM/r},
\end{equation}
\begin{equation}
v_{\phi}(r,\theta ) =v_{\phi}(\theta) \sqrt{GM/r},
\end{equation}
\begin{equation}
c_{s}(r,\theta ) = c_{s}(\theta) \sqrt{GM/r},
\end{equation}
where $\rho(\theta)$ and $c_{s}(\theta)$ are the latitudinal parts of the density and the sound speed, respectively. Also, $v_{r}(\theta)$, $v_{\theta}(\theta)$ and $v_{\phi}(\theta)$ are the latitudinal parts of the components of  velocity in the spherical coordinates.
Upon substituting the above self-similar solutions into the continuity equations and the components of equation of motion, a set of ordinary differential equations is obtained \citep{jiao}:
\begin{displaymath}
2v_{\theta } (\theta ) \frac{d\rho(\theta)}{d\theta } + \rho (\theta ) [ (3-2n) v_{r}(\theta)
\end{displaymath}
\begin{equation}\label{eq:con}
+2 (\cot\theta v_{\theta}(\theta ) + \frac{dv_{\theta} (\theta)}{d\theta}) ]=0,
\end{equation}
\begin{displaymath}
2(n+1) p(\theta ) + \rho (\theta ) [ v_{r}(\theta)^2 + 2 (-1+ v_{\theta}(\theta)^2
\end{displaymath}
\begin{equation}\label{eq:radial}
+v_{\phi}(\theta)^2 - v_{\theta }(\theta ) \frac{dv_{r}(\theta)}{d\theta})]=0,
\end{equation}
\begin{displaymath}
2\frac{dp(\theta)}{d\theta} + \rho (\theta ) [-2\cot\theta v_{\phi}(\theta)^2 + v_{\theta}(\theta) (v_{r}(\theta)
\end{displaymath}
\begin{equation}\label{eq:theta}
+ 2\frac{dv_{\theta}(\theta)}{d\theta})]=0,
\end{equation}
\begin{displaymath}
-2(n-2) \alpha p(\theta) + \rho(\theta) [v_{r}(\theta) v_{\phi}(\theta) + 2 v_{\theta}(\theta) (\cot\theta v_{\phi}(\theta)
\end{displaymath}
\begin{equation}\label{eq:phi}
+\frac{dv_{\phi}(\theta)}{d\theta}]=0,
\end{equation}
If we set $n=3/2$, the self-similar continuity equation (\ref{eq:con}) implies $v_{\theta} =0$. So, we explore other values of $n$ which imply a non-zero value for the latitudinal velocity.

But our energy equation differs from \cite{jiao}, because we are considering thermal conductivity. Although  the  saturated form of the thermal conductivity  is appropriate for ADAFs, \cite{tanaka} provided justifications for   a standard diffusive operator   for  thermal conduction. However, they neglected the latitudinal component of the velocity. Following a similar approach, we also use a standard form of thermal conduction, in which the heat flux depends linearly on the local temperature gradient and moreover the radial dependence of the thermal conductivity coefficient $\lambda $ is assumed to be a power-law function of radius to preserve self-similarity, i.e. $\lambda (r) = \lambda_0 r^{1/2 - n}$. Thus, energy equation becomes
\begin{displaymath}
2\rho (\theta) v_{\theta}(\theta ) \frac{dp(\theta )}{d\theta } +  p(\theta ) \{ \rho(\theta ) [2(n\gamma - n-1)v_{r}(\theta) - 3\alpha f (\gamma -1)
\end{displaymath}
\begin{displaymath}
\times v_{\phi} (\theta)] - 2\gamma v_{\theta }(\theta )\frac{d\rho (\theta )}{d\theta } \} - 2 \lambda_{0} (\gamma -1) \rho (\theta )
\end{displaymath}
\begin{equation}\label{eq:energy}
\times [ (n-\frac{1}{2}) c_{s}(\theta)^2 + \frac{1}{\sin\theta } \frac{d}{d\theta}(\sin\theta \frac{d c_{s}(\theta)^2}{d\theta}) ] =0.
\end{equation}

Equations (\ref{eq:con}), (\ref{eq:radial}), (\ref{eq:theta}), (\ref{eq:phi}) and (\ref{eq:energy}) are our main equation to be solved numerically subject to a set of appropriate boundary conditions. We have five unknown variables, i.e. $\rho(\theta)$, $c_{s}(\theta)$, $v_{r}(\theta)$, $v_{\theta}(\theta)$ and $v_{\phi}(\theta)$. Obviously, the equatorial plane is defined by $\theta = \pi/2$. Assuming that flow is symmetric with respect to the equatorial plane, we have
\begin{equation}
\frac{d\rho}{d\theta}|_{\theta = \pi/2} = \frac{dv_r}{d\theta}|_{\theta=\pi/2}=\frac{dv_{\phi}}{d\theta}|_{\theta=\pi/2}=\frac{dc_s}{d\theta}|_{\theta=\pi/2}=0,
\end{equation}
and $v_{\theta}(\pi/2)=0$. Moreover, we set $\rho(\pi /2) =1$ because we can simply scale the density by the accretion rate. Without thermal conduction, these boundary conditions are sufficient to start numerical integration from the equator to the pole, because  viscous stress tensor is not considered in the radial and the latitudinal components of equation of motion \citep{jiao}. But as we show below, when thermal conductivity is considered we need another physical constraint.
\begin{figure}
\epsfig{figure=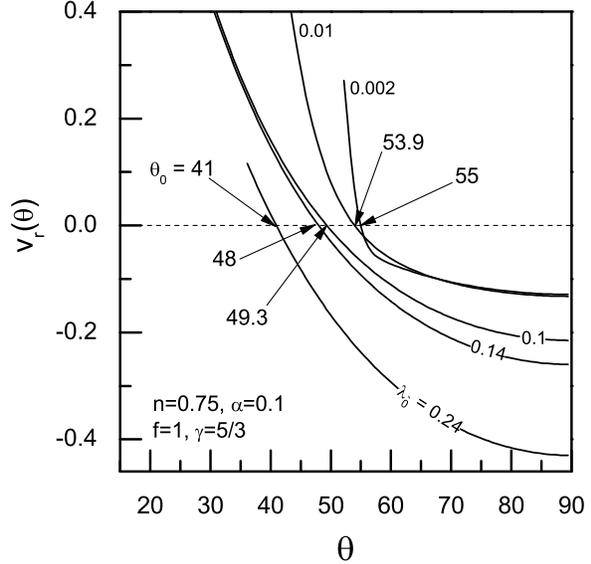,angle=0,scale=0.55}
\caption{Latitudinal profile of the radial velocity for $n=0.75$, $\alpha=0.1$, $f=1$ and $\gamma=5/3$. Each curve is labeled by the corresponding value of the conductivity coefficient $\lambda_0$. The opening angle $\theta_0$ is also marked by arrows.}
\label{fig:f1}
\end{figure}

Substituting the above boundary conditions into equations (\ref{eq:con})-(\ref{eq:energy}), the following equations at the equator are obtained:
\begin{equation}\label{eq:al1}
2(n+1) c_s (\pi/2)^2 + v_r(\pi/2)^2+2(-1+v_{\phi}(\pi/2)^2)=0,
\end{equation}
\begin{equation}\label{eq:al2}
-2(n-2)\alpha c_s (\pi/2)^2 + v_{r}(\pi /2) v_{\phi}(\pi /2)=0,
\end{equation}
\begin{displaymath}
c_s (\pi/2)^2 [2(n\gamma - n-1)v_r (\pi/2) -3\alpha f (\gamma -1) v_{\phi} (\pi/2)]
\end{displaymath}
\begin{equation}\label{eq:al3}
-2\lambda_0 (\gamma -1) [(n-\frac{1}{2})c_{s}(\pi/2)^2 + \frac{d^2 c_s (\theta)^2}{d\theta^2}|_{\theta =\pi/2}]=0,
\end{equation}
If we set $\lambda_0 =0$, the above algebraic equations uniquely give us $v_r(\pi/2)$, $v_{\phi}(\pi/2)$ and $c_{s}(\pi /2)$. Thus, all physical variables at the equator become known and the equations could be solved numerically as has been done in \cite{jiao}. But for a nonzero $\lambda_0$, there is another unknown, i.e. $(d^2 c_s (\theta)^2/d\theta^2)_{\theta=\pi/2}$. It means at the equator, we have three algebraic equations and four unknowns. As long as there is not another physical constraint, one of these unknowns could be taken as a free parameter, say $v_{r}(\pi/2)$. In fact, \cite{Xu} considered the radial velocity at the equator  as a free parameter and constructed their ADAF solutions (without thermal conductivity) for a given  $v_r (\pi/2)$. Instead of following this approach, we construct our solutions subject to a physical boundary condition which was introduced by \cite{Xue}.

\begin{figure}
\vspace{-65pt}
\epsfig{figure=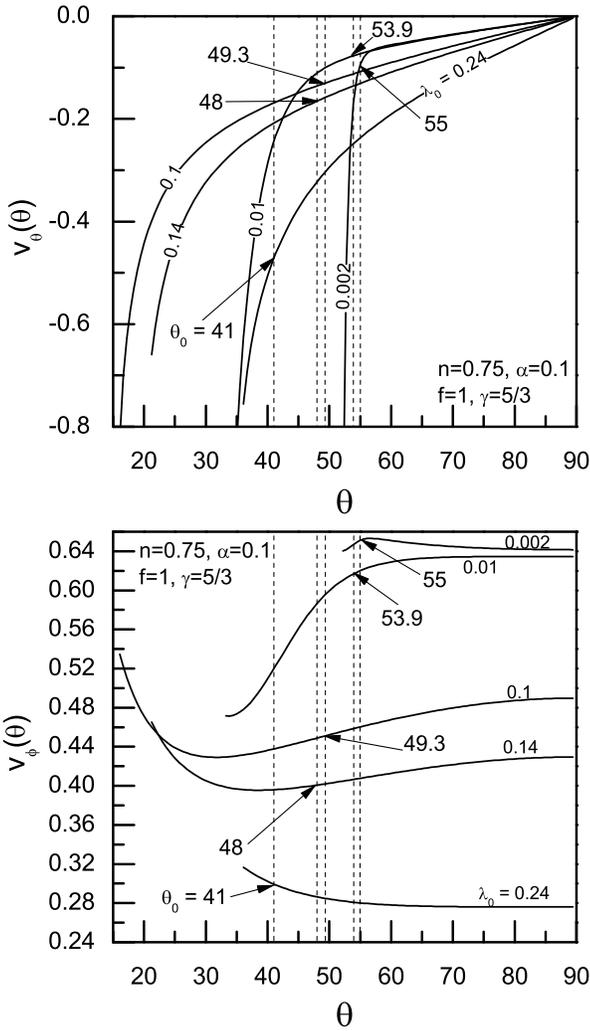,angle=0,scale=0.55}
\caption{Latitudinal profile of the meridional  velocity (top) and the rotational velocity (bottom) for $n=0.75$, $\alpha=0.1$, $f=1$ and $\gamma=5/3$. Corresponding to each curve a vertical dashed line is drawn to separate inflow and outflow regions and the opening angle $\theta_0$ is marked by arrow.}
\label{fig:f2}
\end{figure}

\begin{figure}
\vspace{-65pt}
\epsfig{figure=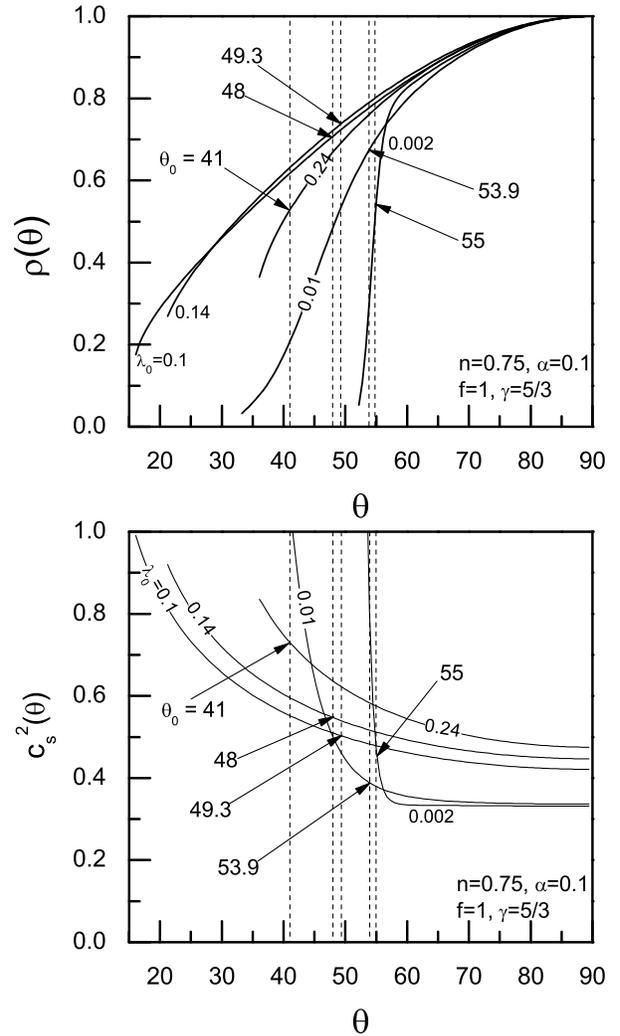,angle=0,scale=0.55}
\caption{Profiles of the density (top) and the sound speed (bottom) are shown. Input parameters are the same as Figures \ref{fig:f1} and \ref{fig:f2}.}
\label{fig:f3}
\end{figure}

In contrary to the solutions in which latitudinal velocity is assumed to be zero \citep[e.g.,][]{narayan95,tanaka},  solutions with nonzero $v_\theta$ do not extend to the pole \citep{Xue,jiao}. According to the solutions, the flow displays  inflow near  the equator and at a certain angle $\theta_0$,  radial velocity becomes zero so that for the regions with $\theta < \theta_0 $ the radial velocity is positive down to the angle $\theta_b$ ($\theta_b < \theta_0$). Numerical integration can not be extended to the regions with $\theta <\theta_b$ because either the density becomes negative which is obviously unphysical or emergence of a singularity. Thus,  inflowing  region extends from equator to $\theta_0$ and beyond which we have outflows. We also found a similar behavior in the presence of thermal conductivity. We can impose a physical constraint at the opening angle $\theta_0$. Since the scale height of the disc is $c_s / \Omega_K$ and it should be equal to the height of the disc, we have $c_s / \Omega_K \cong (\pi/2 - \theta_0)r$ or
\begin{equation}\label{eq:outflow}
c_{s}(\theta_0 )=\pi/2 - \theta_0.
\end{equation}
In other words, this condition implies the height of the disc is proportional to the temperature of outflow. \cite{Xue} constructed their solutions based on this condition, but they prescribed the opening angle $\theta_0$. Although we use this outflow condition to obtain a unique solution for a given set of the input parameters, we do not prescribe the opening angle but it is obtained self-consistently from our numerical integration. Our procedure  is as follows. We first guess the radial velocity at the equator, i.e. $v_r (\pi/2)$, and then $c_s (\pi/2)$ and $v_\phi (\pi/2)$ are obtained from equations (\ref{eq:al1}) and (\ref{eq:al2}). Having all the needed physical variables at the equator and their derivatives, we can integrate equations (\ref{eq:con})-(\ref{eq:energy}) numerically to determine the angle at which the radial velocity becomes zero, i.e. $\theta_0$. Now, we can check if the outflow condition (\ref{eq:outflow}) at the obtained $\theta_0$ is satisfied. If not, we can choose another  $v_r (\pi/2)$ until the outflow condition is satisfied.

Figures \ref{fig:f1}, \ref{fig:f2} and \ref{fig:f3} show the latitudinal profiles of flow dynamical quantities for $n=0.75$, $\alpha=0.1$, $f=1$ and $\gamma=5/3$. In Figure \ref{fig:f1}, the latitudinal dependence of the radial velocity is shown. Each curve is labeled by its value of thermal conductivity coefficient $\lambda_0$. There is an inflow at the equator. But the radial velocity gradually decreases as we move toward the pole so that at a certain angle $\theta_0$ it becomes zero and for $\theta < \theta_0$ it is positive. A horizontal dashed line denotes the transition between the inflow and the outflow regions. Corresponding to each curve, the opening angle $\theta_0$ is marked by an arrow as well. We could not extend numerical integration to the angles less than a certain angle $\theta_b$ where $\theta_b < \theta_0$. As mentioned earlier it has two reasons, either the density becomes negative at $\theta_b$ or a singularity emerges at this angle where the latitudinal velocity becomes equal to the latitudinal part of the sound speed. Therefore, all the solutions are truncated at certain angles. A similar trend has also been observed by \cite{jiao} for the explored configurations without thermal conductivity. As the value of thermal conductivity coefficient $\lambda_0$ is increased, the radial inflow at the equator becomes larger. Also, the opening angle $\theta_0$ reduces with increasing $\lambda_0$. For values of $\lambda_0$ less than $0.002$, we found very little differences between the numerically obtained solutions and are not shown for this reason. Moreover, for values of $\lambda_0$ larger than $0.24$, we found that truncation angle $\theta_b$ appears before the radial velocity becomes zero and so, numerical integration could not be extended to the angles less than $\theta_b$ because of singularity at this angle.

Latitudinal profiles of the meridional and the rotational velocities are shown in Figure \ref{fig:f2}. Vertical dashed lines show transitions between inflowing and outflowing regions for each particular case. Meridional velocity increases as thermal conduction coefficient $\lambda_0$ increases so that with a larger meridional velocity transition to the outflow region occurs. On the other hand, the system rotates more slowly with increasing $\lambda_0$ because the flow is more pressure supported.

Figure \ref{fig:f3} shows latitudinal profiles of the density and the sound speed. As one of our boundary conditions, we normalized density at the equator to unity. The flow approaches spherical symmetry and becomes hotter with increasing $\lambda_0$. Based on the energy consideration, we can explain behavior of the physical quantities which are shown in Figures 2 and 3. In our model, thermal conductivity provides an extra energy of source of heating in the latitudinal direction. The radial conduction term is tied to a fixed value set by self-similarity.  As the coefficient of thermal conduction increases, more energy is transported by the thermal conduction and the flow becomes hotter (bottom plot of Figure 3). Thus, a flow with thermal conduction becomes more pressure supported in comparison to the same flow without thermal conduction. This fact implies a more slowly rotating flow as thermal conductivity increases (bottom plot of Figure 2). Since the thickness of the inflowing region is proportional to the sound speed (or temperature), the flow tends to a more spherical configuration  as it becomes hotter with the increase of thermal conduction (top plot of Figure 3).

\begin{figure}
\epsfig{figure=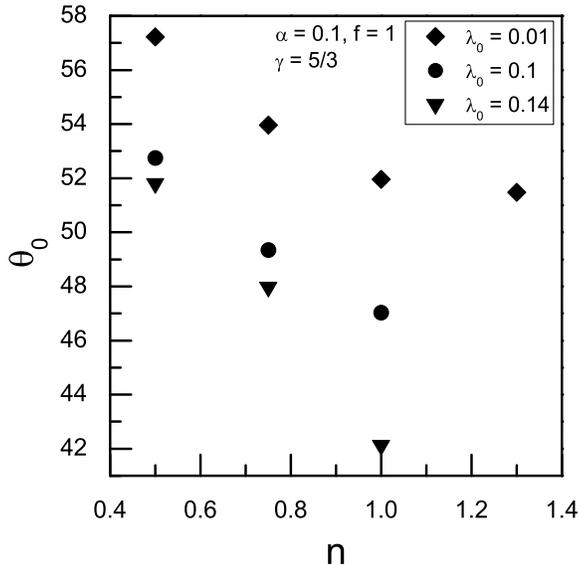,angle=0,scale=0.55}
\caption{Opening angle $\theta_0$ versus the parameter $n$ with $\alpha=0.1$, $f=1$ and $\gamma=5/3$ and various degrees of conduction as measured by $\lambda_0$.}
\label{fig:f4}
\end{figure}
Dependence of the opening angle of the outflowing region on the coefficient $\lambda_0$ and $n$ is interesting.  Figure \ref{fig:f4} shows opening angle as a function of $n$ for $\alpha=0.1$, $f=1$ and $\gamma=5/3$ and different values of $\lambda_0$. As the value of $n$ increases, the opening angle reduces for a fixed coefficient $\lambda_0$. But irrespective of the value of $n$, the opening angle reduces with increasing thermal conductivity coefficient.  We also explored other values for the input parameters and the results are qualitatively similar to what have been presented so far. Moreover, our goal was to investigate possible effects of thermal conductivity, and the roles of the other input parameters have been reported before \citep{jiao}.

\section{Conclusions}
We studied two-dimensional structure of an ADAF with bipolar outflows in the presence of thermal conduction. In most of the previous works, height-integrated hydrodynamical equations have been used to study the dynamical effects of thermal conduction. These illustrative models are useful, though role of thermal conduction in launching outflows and their properties like opening angle could not be studied based on these models. Moreover, ADAFs are generally hot and geometrically thick  and it is more convenient to study their structure in the spherical coordinates. Previous semi-analytical studies of ADAFs in spherical coordinates either included all three components of the gas velocity but did not consider thermal conduction \citep{Xu,Xue,jiao} or only radial and rotational components of velocity are considered including  energy transport by the thermal conduction \citep{tanaka}. In this regard, our study differs from previous studies because not only we considered all three components of the gas velocity, but  thermal conductivity in the energy equation is also included. This enabled us to study some of the features of our advective flows, in particular possible role of thermal conduction in properties of outflows.

In our self-similar solutions, energy transport by the thermal conductivity is self-similar in the radial direction, but in the latitudinal direction it is determined from the dynamical equations. Thus, as thermal conduction coefficient increases, the flow becomes hotter due to enhancement of the transported energy by thermal conduction (see bottom plot of Figure 3). Thus, transition from inflowing region to the outflowing region occurs at a larger temperature with the increase of conductivity. This point along with our imposed boundary condition (i.e., equation (18)) implies a smaller $\theta_0$ when thermal conduction increases. Therefore, such a latitudinal energy transport by the thermal conduction provides an extra energy source for launching outflows in a hotter flow comparing to the same flow without thermal conduction.
Interestingly, it was found the radial component of the gas velocity changes its sign at a certain angle which implies both inflow and outflow exist in an advective flow with thermal conduction. Such a trend has already been reported in the flows without thermal conduction \citep{jiao}. But the size of the outflowing and inflowing regions are significantly modified because of the existence of thermal conductivity according to our solutions. As thermal conduction increases, the opening angle of the outflow decreases which implies a more collimated outflow.  In other words, thermal conduction contributes to the collimation and emergence of narrower outflows. However, properties of outflow are not described based on our solutions. The only point we can mention is that the size of outflowing region decreases with the increase of thermal conductivity. Also, we can state that at the transition angle $\theta_0$, the flow has a larger latitudinal velocity when thermal conductivity increases (see top plot of Figure 2). Our similarity solutions are not valid within the outflowing region. Thus, it is not possible to determine the strength and the speed of outflows in the presence of thermal conduction just based on our similarity solutions. Our parameter space survey covered other ranges of the input parameters and the results are similar to what we are just describing and have been shown by our illustrative Figures.

Although our similarity solutions display both inflowing and outflowing regions, self-similarity is broken in the outflowing region. Emergence of bipolar outflows and their opening angles could be investigated using similarity solutions. It would be interesting to relax self-similar assumption and study global solutions in the presence of thermal conduction. In doing so, energy transport by thermal conductivity is not radially self-similar and it affects both the polar and the radial temperature profiles. Semi-analytical similarity solutions, however, provide a physical insight of what we can expect from such a global and more reasonable analysis.

\section*{Acknowledgments}
We are grateful to the anonymous referee whose detailed and careful comments helped to improve the quality of this paper.

\bibliographystyle{mn2e}
\bibliography{referenceKH}


%
%
%
\end{document}